\documentclass[12pt]{article}
\usepackage{amssymb}
\usepackage{amsmath}
\usepackage[space]{cite}
\usepackage{algorithm}
\usepackage[noend]{algpseudocode}
\usepackage[normalem]{ulem}
\usepackage{url}
\usepackage{graphicx}

\numberwithin{equation}{section}

\newcommand{\be}{\begin{eqnarray}}
\newcommand{\ee}{\end{eqnarray}}
\newcommand{\non}{\nonumber}
\newcommand{\id}{\mathbb{I}}

\newcommand{\diag}{\mathop{\rm diag}\nolimits}

\usepackage[usenames,dvipsnames]{color}

\newcommand{\cut}[1]{\ifmmode\text{\textcolor{red}{\sout{\ensuremath{#1}}}}\else\textcolor{red}{\sout{#1}}\fi}

\begin{document}

\begin{titlepage}
\strut\hfill UMTG--313
\vspace{.5in}
\begin{center}

{\LARGE Bethe states on a quantum computer:\\[0.2in]
success probability and correlation functions}\\
\vspace{1in}
\large Wen Li, Mert Okyay and Rafael I. Nepomechie\\[0.2in] 
\large Physics Department, P.O. Box 248046, University of Miami\\[0.2in]  
\large Coral Gables, FL 33124 USA\\
\end{center}
{\let\thefootnote\relax\footnote{{{\tt  
wxl386@miami.edu, mokyay@miami.edu, nepomechie@miami.edu}}}}

\vspace{.5in}

\begin{abstract}
A probabilistic algorithm for preparing Bethe eigenstates of the
spin-1/2 Heisenberg spin chain on a quantum computer has recently been
found.  We derive an exact formula for the success probability of this
algorithm in terms of the Gaudin determinant, and we study its
large-length limit.  We demonstrate the feasibility of computing
antiferromagnetic ground-state spin-spin correlation functions for
short chains.  However, the success probability decreases
exponentially with the chain length, which precludes the computation
of these correlation functions for chains of moderate length.  Some
conjectures for estimates of the Gaudin determinant are noted in an
appendix.
\end{abstract}

\end{titlepage}

\setcounter{footnote}{0}

\section{Introduction}\label{sec:intro}

The Hamiltonian of the closed isotropic spin-1/2 Heisenberg (or XXX) quantum 
spin chain of length $L$ with periodic boundary conditions is given 
by 
\begin{equation}
{\cal H} = \tfrac{1}{2}\sum_{n=0}^{L-1} \left(\vec\sigma_{n} \cdot 
\vec\sigma_{n+1} -1 \right)\,, \qquad \vec\sigma_{L} 
= \vec\sigma_{0} \,,
\label{Hamiltonian}
\end{equation}
where $\vec\sigma_{n} \cdot \vec\sigma_{n+1} = \sigma^{x}_{n}\, \sigma^{x}_{n+1} + \sigma^{y}_{n}\, \sigma^{y}_{n+1}  
+ \sigma^{z}_{n}\, \sigma^{z}_{n+1}$, and
as usual $\sigma^{x}_{n}\,, \sigma^{y}_{n}\,, \sigma^{z}_{n}$
are Pauli matrices at site $n$.	This model was solved by Bethe 
\cite{Bethe:1931hc} using an approach now known as coordinate Bethe 
ansatz. Roughly speaking, the exact eigenstates (``Bethe states'') of 
the Hamiltonian \eqref{Hamiltonian}
are given by $M$-particle states (where $M=0, 1, \ldots, L/2$), which are expressed in terms of $M$ 
quasi-momenta (``Bethe roots''), which in turn are solutions of a 
system of $M$ equations (``Bethe equations''). This  
remarkable solution is made possible by the fact that this model -- 
among infinitely many others -- is 
quantum integrable (see e.g. \cite{Faddeev:1981ft, Gaudin:1983} and references therein).

An algorithm for preparing these Bethe states (corresponding to real Bethe 
roots) on a quantum computer has recently 
been found \cite{VanDyke:2021kvq}. We henceforth refer here to this 
algorithm as the ``Bethe algorithm'', and to the corresponding 
quantum circuit as the ``Bethe circuit''. The Bethe algorithm was actually 
formulated for the more general case 
of the anisotropic (or XXZ) quantum spin chain, with anisotropy 
parameter $\Delta$. However, for clarity, we focus here on the isotropic case 
$\Delta=1$.

An important feature of the Bethe algorithm is that -- like the 
algorithms in \cite{Childs:2012, Berry:2015} -- it is {\em 
probabilistic}.\footnote{A deterministic construction of Bethe states 
seems difficult \cite{Nepomechie:2021bethe}.} The success 
probability (that is, the probability of generating a desired Bethe 
state) was determined in \cite{VanDyke:2021kvq} ``experimentally'' by running the 
algorithm on the IBM Qiskit statevector simulator. One of our main results is a 
simple exact formula for the success probability in terms of the  so-called 
Gaudin determinant \cite{Gaudin:1983, Gaudin:1981cyg, Korepin:1982gg}, 
see Eq. \eqref{successprobab} below. 
We also argue that, for large $L$ and fixed number of 
Bethe roots $M$, the success probability approaches $1/M!$ \eqref{limit1}.

The Hamiltonian \eqref{Hamiltonian} describes an {\em antiferromagnetic} 
spin chain (notice that the coefficient of $\vec\sigma_{n} \cdot \vec\sigma_{n+1}$ 
is positive), and therefore it has a nontrivial N\'eel-like ground state.
Many results for this model's spin-spin correlation functions are 
already known, mostly for small and large values of $L$ (see e.g. 
\cite{Hulthen:1938, Luther:1975wr, Takahashi:1977, Lin:1991, Hallberg:1995, 
Kitanine:2000srg, Kitanine:2002mn, Lukyanov:2002fg, Sakai:2003pc, 
Boos:2004sk, Boos:2005vu, Korepin:2006jvi} and 
references therein). Much less is known for intermediate size ($1 \ll 
L \ll \infty$), see however \cite{Caux:2005a, Caux:2005b, Caux:2009} 
for results obtained using the ABACUS algorithm; 
and one can ask whether the Bethe algorithm could be used to compute 
such correlation functions, once appropriate hardware 
becomes available.\footnote{The use of quantum computers to
compute correlation functions has been considered in e.g. 
\cite{Somma:2002, Wecker:2015}.}
However, due to the probabilistic nature of this 
algorithm, it is not obvious how to set up such computations. We 
describe a way of measuring the correlation functions, and we estimate 
the number of shots needed for a given error. These analyses are 
supported by numerical simulations for small values of $L$. We find 
that the success probability decreases exponentially with $L$, which precludes the
computation of these correlation functions for moderate values of $L$,
as anticipated in \cite{VanDyke:2021kvq}.

The outline of the remainder of this paper is as follows. In Sec. 
\ref{sec:basics}, we briefly review the coordinate Bethe ansatz 
solution of the model, and the Bethe circuit for preparing Bethe 
states on a quantum computer. In Sec. \ref{sec:sucprob} we derive an 
exact expression for the probability that the Bethe circuit 
successfully prepares a given Bethe state, and we 
study its large-$L$ limit. In Sec. \ref{sec:corr} we 
investigate the application of the Bethe circuit to computing the model's
spin-spin correlation functions. Sec. \ref{sec:end} 
contains a brief discussion of our results. In appendix \ref{sec:conjectures}
we note some conjectures for estimates of the Gaudin 
determinant, which arose from our study of the success probability,
which may be of independent interest.

\section{Bethe basics}\label{sec:basics}

We briefly review here the coordinate Bethe ansatz solution of the 
model \eqref{Hamiltonian}, and the Bethe circuit 
\cite{VanDyke:2021kvq} for preparing Bethe states on a quantum 
computer.

\subsection{Coordinate Bethe ansatz}\label{sec:CBA}

Let us assume that $\{ k_{0}, \ldots, k_{M-1}\}$ are 
pairwise distinct and satisfy the Bethe equations
\begin{equation}
e^{i k_{j} L} = \prod_{l=0; l \ne j}^{M-1} S(k_{j}, k_{l})\,, \qquad j 
= 0, \ldots, M-1 \,, 
\label{BE}
\end{equation}
where
\begin{equation}
S(k_{j}, k_{l}) = \frac{u(k_{j}) - u(k_{l}) + i}
{u(k_{j}) - u(k_{l}) - i}\,, \qquad 
u(k)=\frac{1}{2}\cot\left(\frac{k}{2}\right) \,,
\end{equation}
and $M=0, 1, \ldots, L/2$.
For real $k$'s, 
\begin{equation}
S(k_{j}, k_{l}) = -e^{i \Theta(k_{j}, k_{l})}\,, \qquad 	\Theta(k_{j}, 
k_{l}) = 2 
\arctan\left[\frac{\sin(\frac{1}{2}(k_{j}-k_{l}))}
{\cos(\frac{1}{2}(k_{j}-k_{l}))-\cos(\frac{1}{2}(k_{j}+k_{l}))}\right] \,.
\end{equation}

The corresponding Bethe state is given by
\begin{equation}
|\psi \rangle = \sum_{0 \le x_{0} < x_{1} < \ldots < x_{M-1} \le L-1} 
f(x_{0}, \ldots, x_{M-1}) |x_{0}, \ldots, x_{M-1} \rangle \,,
\label{Bethestate}
\end{equation}
where 
\begin{equation}
|x_{0}, \ldots, x_{M-1} \rangle = \sigma^{-}_{x_{0}} \ldots 
\sigma^{-}_{x_{M-1}} |0 \ldots 0 \rangle\,,
\end{equation}
with $\sigma^{-}_{n}=\frac{1}{2}(\sigma^{x}_{n}- i \sigma^{y}_{n})$ is the 
spin-lowering operator at site $n$, and $|0 \ldots 0 \rangle$ is the ferromagnetic ground state
(i.e., the reference state with all $L$ spins in the up-state $|0\rangle = {1\choose 0}$.) 
The wave function $f(x_{0}, \ldots, x_{M-1})$ is given by
\be
f(x_{0}, \ldots, x_{M-1}) = \sum_{P} \varepsilon_{P}\, A_{P}\, 
e^{i \sum_{j=0}^{M-1} k_{P(j)} x_{j}}\,,
\label{wavefunction}
\ee 
where the sum is over all $M!$ permutations 
$P: \{ 0, \ldots, M-1\} \xrightarrow{\rm{\tiny bijection}} \{ 0, \ldots, M-1\}$,
and $\varepsilon_{P}$ denotes the signature of the permutation.
Moreover, the amplitudes $A_{P}$ satisfy 
\begin{equation}
\frac{A_{P}}{A_{P'}} = -S(k_{P(l)}, k_{P'(l)})\,, \qquad A_{I} = 1\,,
\end{equation}
where $P$ and $P'$ are permutations that differ by a single 
transposition between adjacent elements, $P'(l)=P(l+1)$ and 
$P'(l+1)=P(l)$ for some $l \in \{0,\ldots, M-1\}$, 
and $I$ is the identity permutation.

The fact that the $k$'s satisfy the Bethe equations 
\eqref{BE} implies that the Bethe state \eqref{Bethestate} is ``on shell''; i.e.,
it is an eigenstate of the Hamiltonian \eqref{Hamiltonian} 
\begin{equation}
{\cal H}\, |\psi \rangle = E\, |\psi \rangle\,, \qquad E = 
-\sum_{j=0}^{M-1} 4 \sin^{2}\left(\frac{k_{j}}{2}\right) \,.
\label{energy}
\end{equation}
Throughout this paper, all Bethe states should be understood to be on shell.
	
\subsection{The Bethe circuit}\label{sec:circuit}

\begin{figure}[htb]
	\centering
	\includegraphics[width=0.8\hsize]{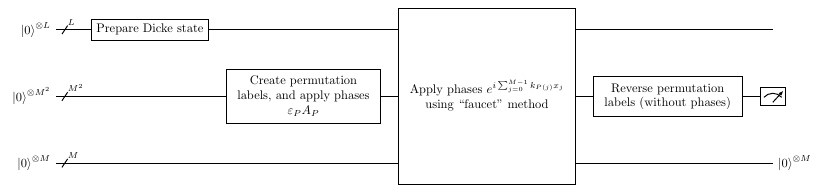}
	\caption{Schematic diagram of the Bethe circuit}
	\label{fig:circuit}
\end{figure}

The Bethe circuit \cite{VanDyke:2021kvq} for preparing on a quantum 
computer the state \eqref{Bethestate} with all $k$'s real
is depicted schematically in Fig \ref{fig:circuit}. 
Note that there are $L$ ``system'' qubits, $M^{2}$ ``permutation-label'' qubits, and $M$ 
``faucet'' qubits.

This circuit proceeds by the following 5 main steps (see 
\cite{VanDyke:2021kvq} for details):
\begin{enumerate}
	
\item Prepare the system qubits in the so-called Dicke state
\begin{equation}
\frac{1}{\sqrt{{L\choose M}}}\sum_{0 \le x_{0} < \ldots < x_{M-1} \le L-1}  
|x_{0}, \ldots, x_{M-1} \rangle \,.
\label{Dicke}
\end{equation}

\item Prepare the permutation-label qubits in the state
\begin{equation}
\frac{1}{\sqrt{M!}}\sum_{P} \varepsilon_{P}\, A_{P} |P\rangle \,,
\label{PLstate}
\end{equation}
where the states $|P\rangle$ store the permutations by 
means of ``one-hot encoding''.

\item Apply the phases 
$e^{i \sum_{j=0}^{M-1} k_{P(j)} x_{j}}$ using the ``faucet'' method. 
At the end of this step, the faucet qubits return to their original 
state $|0\rangle^{\otimes M}$.

\item Reverse step 2, except without the phases 
$\varepsilon_{P}\, A_{P}$. 

\item Measure the permutation-label qubits, with success (namely, the system 
qubits are in a state proportional to the Bethe state 
\eqref{Bethestate}) on $|0\ldots 0 \rangle$.

\end{enumerate}

\section{Success probability}\label{sec:sucprob}

We now compute the success probability of the Bethe circuit.

Just prior to the measurements (i.e. at the end of step 4), the Bethe circuit 
brings the quantum computer to the state
\begin{equation}
|\Psi \rangle = \frac{1}{\sqrt{{L\choose M}}} \frac{1}{M!} 
|0\ldots 0\rangle | \psi \rangle + \ldots \,, \qquad \langle \Psi 
|\Psi\rangle = 1 \,,
\label{statePsi}
\end{equation}
where $| \psi \rangle$ is the unnormalized state \eqref{Bethestate}, 
and the ellipsis denotes additional terms that are 
orthogonal to the first.
The factor $\frac{1}{\sqrt{{L\choose M}}}$ comes from the operator 
that creates the Dicke state \eqref{Dicke}
at step 1, and the factor $\frac{1}{M!}=\left(\frac{1}{\sqrt{M!}}\right)^{2}$
comes from the operator 
that creates the permutation-label state \eqref{PLstate} and its 
inverse, see steps 2 and 4 above.

For later convenience, let us define the state $|\phi \rangle$ by the following rescaling of $| \psi \rangle$  
\begin{equation}
|\phi \rangle := \left[\prod_{0\le j < l \le M-1} (j,l) \right] |\psi 
\rangle \,,
\label{phi}
\end{equation}
where we have introduced the notation \cite{Gaudin:1981cyg}
\begin{equation}
(j,l) := 2 - e^{-i k_{j}} - e^{i k_{l}} \,.
\end{equation}
In terms of this notation, the state \eqref{statePsi} is given by
\begin{align}
|\Psi \rangle &= \frac{1}{\sqrt{{L\choose M}}} \frac{1}{M!} 
\frac{1}{\left[\prod_{0\le j < l \le M-1} (j,l) \right]}
|0\ldots 0\rangle | \phi \rangle + \ldots \,, \non \\
&= \alpha |0\ldots 0\rangle | \tilde{\phi} \rangle + \ldots
\label{statePsifinal}
\end{align}
where $| \tilde{\phi} \rangle$ is the normalized state
\begin{equation}
|\tilde{\phi} \rangle := \frac{| \phi \rangle}{\sqrt{\langle \phi | 
\phi \rangle}} \,, \qquad \langle \tilde{\phi} | \tilde{\phi} \rangle 
= 1\,,
\end{equation}
and $\alpha$ is defined by
\begin{equation}
\alpha := \frac{1}{\sqrt{{L\choose M}}} \frac{1}{M!} 
\frac{1}{\left[\prod_{0\le j < l \le M-1} (j,l) \right]} \sqrt{\langle \phi | 
\phi \rangle} \,.
\end{equation}

The probability that all the ancillary qubits are in the state $|0\ldots 
0\rangle$, and that the Bethe state $| \tilde{\phi} \rangle$ has 
therefore been successfully prepared, is $|\alpha|^{2}$.
Since all the Bethe roots $\{ k_{i} \}$ are real, $(j,l)^{*} = 
(l,j)$. Hence, the success probability is given by
\begin{equation}
|\alpha|^{2} = \frac{1}{{L\choose M}} \frac{1}{\left(M!\right)^{2}} 
\frac{1}{\left[\prod_{0\le j < l \le M-1} (j,l)(l,j) \right]} 
\langle \phi | \phi \rangle \,.
\label{alpha2}
\end{equation}

We observe that the Bethe wavefunction corresponding to the 
unnormalized state $|\phi\rangle$ \eqref{phi} is normalized in the 
same way as in  \cite{Gaudin:1981cyg}. (Indeed, $|\phi\rangle$
also has the form \eqref{wavefunction}, 
but with $A_{I} \ne 1$, and with $A_{P}$ the same (up to an irrelevant 
phase) as in \cite{Gaudin:1981cyg}.)
The squared norm $\langle \phi | \phi \rangle$ is therefore given by 
\cite{Gaudin:1981cyg}\footnote{There is an extra factor $N!$ (corresponding 
to $M!$ in our notation) in Eq. (24) of \cite{Gaudin:1981cyg} that 
is absent in our conventions.}
\begin{equation}
\langle \phi | \phi \rangle = \left[\prod_{0\le j < l \le M-1} 
(j,l)(l,j) \right]\, \det G\,,
\label{squarednorm}
\end{equation}
where $G$ is the so-called Gaudin matrix, which is an $M \times M$ 
matrix whose components are given by 
\begin{equation}
G_{m,n} = \delta_{m,n} \left[L - \sum_{l=0}^{M-1} 
\frac{4\left(1- \cos k_{l}\right)}{(n,l)\, (l,n)} \right] + 
\frac{4\left(1- \cos k_{m}\right)}{(n,m)\, (m,n)}\,, \qquad m, n \in 
\{0, \ldots, M-1\} \,.
\label{Gaudinmatrix}
\end{equation}	
We conclude from \eqref{alpha2} and \eqref{squarednorm}
that the success probability is given by\footnote{For the XXZ spin 
chain with anisotropy parameter $\Delta$, the result is the same, except $G$ is then given by \cite{Gaudin:1981cyg} 
\begin{equation}
G_{m,n} = \delta_{m,n} \left[L - \sum_{l=0}^{M-1} 
\frac{4 \Delta\left(\Delta- \cos k_{l}\right)}{(n,l)\, (l,n)} \right] + 
\frac{4\Delta\left(\Delta- \cos k_{m}\right)}{(n,m)\, (m,n)}\,, \qquad m, n \in 
\{0, \ldots, M-1\} \,, \non
\end{equation}	
with 
\begin{equation}
(j,l) := 2\Delta - e^{-i k_{j}} - e^{i k_{l}} \,. \non
\end{equation}
}
\begin{equation}
|\alpha|^{2} = \frac{1}{{L\choose M}} \frac{1}{\left(M!\right)^{2}} \, \det G
= \frac{(L-M)!}{L!\, M!} \det G \,.
\label{successprobab}
\end{equation}

The result \eqref{successprobab} for the success probability is one of
the main results of this paper.  Using this formula, we obtain the
results in Table \ref{table:successprobab}, which coincide with
corresponding results obtained by running the Bethe circuit on the IBM
Qiskit statevector simulator.

\begin{table}[htb]
\centering
\begin{tabular}{|c|c|c|c|}
\hline
$L$ & $M$ &  $k_{0}, \ldots, k_{M-1}$ & $|\alpha|^{2}$\\   
\hline
4 & 2  & $\pm 2\pi/3$ &  0.5 \\
6 & 2  & 1.41951, 2.76928 & 0.463068 \\
6 & 3  & $\pm 1.72277, \pi$ & 0.157232 \\
8 & 4  & $\pm 1.522, \pm 2.63483$ &  0.0361418 \\
\hline
\end{tabular}
\caption{Success probabilities computed using \eqref{successprobab}}\label{table:successprobab}
\end{table}

We can now argue that, for a 
fixed finite number of Bethe roots $M$, the large-$L$ limit of the success 
probability is given by
\begin{equation}
\lim_{L\rightarrow \infty} |\alpha|^{2} = \frac{1}{M!}	\,.
\label{limit1}
\end{equation}
Indeed, in view of the relation \eqref{successprobab}, it suffices 
to show that
\begin{equation}
\lim_{L\rightarrow \infty} \frac{(L-M)!}{L!}\det G = 
1 \,.
\label{limit2}
\end{equation}
To this end, we observe that, for large $L$, the Gaudin matrix 
\eqref{Gaudinmatrix} is (up to terms of order 1) proportional to the 
identity matrix\footnote{This step of the argument is 
admittedly heuristic, since the $k$'s depend on $L$ through the Bethe 
equations \eqref{BE}, and the denominators $(n,l)(l,n)$ can in 
principle be small. It would be desirable to find a rigorous derivation of 
\eqref{GlargeL}, see also Appendix \ref{sec:conjectures}.} 
\begin{equation}
G \sim  \id L \qquad \text{ for   }\quad L\rightarrow \infty\,,
\label{GlargeL}
\end{equation}
which implies that $\det G \sim L^{M}$ (up to terms of order $L^{M-1}$). Hence,
\begin{equation}
\lim_{L\rightarrow \infty} \frac{(L-M)!}{L!}\det G 
= \lim_{L\rightarrow \infty} \frac{(L-M)!}{L!} L^{M} =1 \,,
\end{equation}
as claimed. 

\begin{figure}[htb]
	\centering
	\includegraphics[width=0.5\hsize]{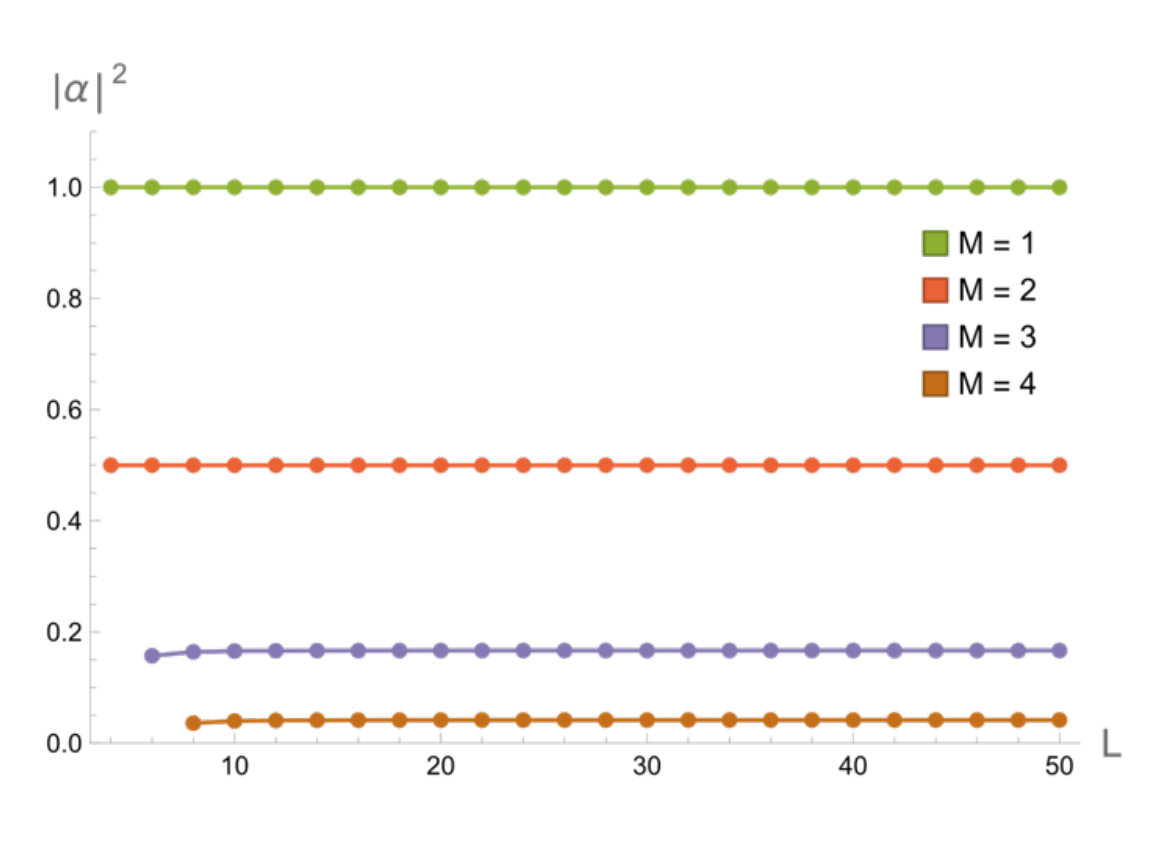}
	\caption{Success probability $|\alpha|^{2}$ for low-energy states 
	with $M=1, 2, 3, 4$ as a function of chain length $L$}
	\label{fig:successprobabM1234}
\end{figure}

We checked the result \eqref{limit1} by numerically studying the
success probability for fixed values of $M$, as a function of $L$. 
However, for given values of $M$ and $L$, there are generally multiple real solutions 
of the Bethe equations. For definiteness, we selected the
ones with {\em lowest energy} \eqref{energy}. For brevity, we 
refer to the corresponding states as ``low-energy states''. For 
the cases $M=1, 2, 3,4$, we plotted the success probabilities 
for these low-energy states as a function of $L$, with $L = {\rm even}$, as shown in Fig.
\ref{fig:successprobabM1234}. This data is clearly consistent with 
the result \eqref{limit1}.

The result \eqref{limit1} supports the 
count estimates in \cite{VanDyke:2021kvq} of the 
non-Clifford gates in the Bethe circuit.

\section{Correlation functions}\label{sec:corr}

For concreteness, we consider here the spin-spin 
correlation functions
\begin{equation}
	\langle \psi_{0}| \sigma^{z}_{0}\, \sigma^{z}_{l}| \psi_{0} \rangle \,, \qquad l = 1, 
	2, \ldots, \frac{L}{2} \,,
\label{corrfunc}
\end{equation}
with $L$ even, where $| \psi_{0} \rangle$ is the normalized ground state of the 
antiferromagnetic Hamiltonian \eqref{Hamiltonian}, which is described 
by $M=\frac{L}{2}$ real Bethe roots (see e.g. 
\cite{Faddeev:1981ft, Gaudin:1983}). As already noted in the 
Introduction, there is an extensive literature on these correlation functions, see e.g. \cite{Hulthen:1938, Luther:1975wr, Takahashi:1977, Lin:1991, Hallberg:1995, 
Kitanine:2000srg, Kitanine:2002mn, Lukyanov:2002fg, Sakai:2003pc, 
Boos:2004sk, Boos:2005vu, Korepin:2006jvi, Caux:2005a, Caux:2005b, Caux:2009} and references therein. 
Our goal is to determine if -- and if so, to what extent -- these 
correlation functions can be computed using a quantum computer.

\subsection{Measuring correlators}

In order to perform shot-based measurements of the expectation values \eqref{corrfunc} on a quantum computer, 
we cannot take advantage of built-in functionality (say, in Qiskit) for computing expectation 
values, since we are preparing the state $| \psi_{0} \rangle$ probabilistically.
Nevertheless, as shown below, we can proceed by simply measuring all the qubits, 
and then appropriately combining the corresponding probabilities, which can be 
approximated from the corresponding counts. 

As a first warm-up exercise, let us consider a quantum computer in 
the 2-qubit state $|\Psi\rangle$ given by
\begin{equation}
|\Psi\rangle = |0\rangle |a\rangle + 
|1\rangle |b\rangle\,, \qquad \langle \Psi |\Psi\rangle =
\langle a|a\rangle + \langle b|b\rangle 
= 1\,,
\label{Psiwarmup}
\end{equation}
where $|a\rangle$ and $|b\rangle$ are (unnormalized) 1-qubit states. 
Suppose that we wish to compute an expectation value in the state 
$|a\rangle$ (rather than the full state $|\Psi\rangle$), for example,
$\langle a| \sigma^{z} |a\rangle/\langle a|a\rangle$. 
According to the Born rule, upon
measuring both qubits of the state $|\Psi\rangle$ in the 
computational basis, $|\Psi\rangle$  is projected to the 
computational basis states $|i_{1}\ i_{0} \rangle$
with probabilities
\begin{equation}
p_{i_{1} i_{0}} = |\langle i_{1}\ i_{0} | \Psi\rangle |^{2} \,, 
\qquad  i_{0}, i_{1} \in \{ 0\,, 1 \} \,,
\end{equation}
which can be approximated from the corresponding counts.
Setting
\begin{equation}
|a\rangle = \sum_{i=0}^{1} a_{i} |i	\rangle = {a_{0}\choose a_{1}} \,,
\end{equation}
and similarly for $|b\rangle$, we see that
\begin{equation}
p_{00} = |a_{0}|^{2}\,, \quad 
p_{01} = |a_{1}|^{2}\,, \quad 
p_{10} = |b_{0}|^{2}\,, \quad 
p_{11} = |b_{1}|^{2}\,.
\end{equation}
We therefore obtain an expression for 
$\langle a| \sigma^{z} |a\rangle/\langle a |a\rangle$ in terms of probabilities $p_{0i}$
\begin{equation}
\frac{\langle a| \sigma^{z} |a\rangle}{\langle a |a\rangle}
= \frac{|a_{0}|^{2} - |a_{1}|^{2}}{|a_{0}|^{2} + |a_{1}|^{2}}
= \frac{p_{00} - p_{01}}{p_{00} + p_{01}} \,.
\label{warmup1}
\end{equation}

As a second warm-up exercise, let us now suppose that the state $|\Psi\rangle$ 
in Eq. \eqref{Psiwarmup} is a 3-qubit state,  $|a\rangle$ and 
$|b\rangle$ now being 2-qubit states; and we now wish to compute the 
2-qubit expectation value $\langle a | \sigma_{0}^{z} \, \sigma_{1}^{z} |a\rangle/\langle 
a |a\rangle$. Setting
\begin{equation}
|a\rangle = \sum_{i_{0}, i_{1}=0}^{1} a_{i_{1} i_{0}} |i_{1}\ i_{0}	\rangle \,,
\end{equation}
we obtain in a similar way an expression for the desired expectation value in terms of the 
probabilities $p_{0 i_{1} i_{0}}$
\begin{align}
\frac{\langle a | \sigma_{0}^{z} \, \sigma_{1}^{z}|a\rangle}
{\langle a |a\rangle}	&= \frac{|a_{00}|^{2} - |a_{01}|^{2} - 
|a_{10}|^{2} + |a_{11}|^{2}}
{|a_{00}|^{2} + |a_{01}|^{2} + |a_{10}|^{2} + |a_{11}|^{2}}  \non \\
&= \frac{p_{000} - p_{001} - p_{010} + p_{011}}
{p_{000} + p_{001} + p_{010} + p_{011}} \,.
\label{warmup2}
\end{align}

Let us now return to the original problem of evaluating the correlators \eqref{corrfunc}. We first
use the Bethe circuit to bring the quantum computer to the state (see Eq. 
\eqref{statePsifinal})
\begin{equation}
|\Psi \rangle = \alpha |0\ldots 0\rangle | \psi_{0} \rangle + \ldots \,,
\end{equation}
and then we simply measure all the qubits. Since the operators $\sigma^{z}_{0}\, 
\sigma^{z}_{l}$ are diagonal, the expectation values in the state 
$|\psi_{0} \rangle$ are given in terms of the
probabilities $p_{0\ldots 0\, i_{L-1} \ldots i_{0}}$ by 
\begin{equation}
\langle \psi_{0}| \sigma^{z}_{0}\, \sigma^{z}_{l}| \psi_{0} \rangle = 
\frac{\sum_{i_{0}, \ldots, i_{L-1}=0}^{1} 
c^{(l)}_{i_{L-1} \ldots i_{0}}\, p_{0\ldots 0\, i_{L-1} \ldots i_{0}}}
{\sum_{i_{0}, \ldots, i_{L-1}=0}^{1} p_{0\ldots 0\, i_{L-1} 
\ldots i_{0}}} \,,
\label{corrmeas}
\end{equation}
where the coefficients $c^{(l)}_{i_{L-1} \ldots i_{0}}$
(with $i_{k} \in \{ 0\,, 1 \}$)
are either 1 or -1, as in \eqref{warmup1} and \eqref{warmup2}. 

The problem therefore reduces to determining the correct signs 
$c^{(l)}_{i_{L-1} \ldots i_{0}}$. To this end, let us introduce the 
diagonal $2^{L} \times 2^{L}$ matrices $D[n]$ whose diagonal elements 
are either 1 or -1, and which alternate every $n$ elements. As 
examples, for $L=2$:
\begin{align}
D[1] &= \diag\left(1,-1,1,-1\right) \non \\
D[2] &= \diag\left(1,1,-1,-1\right) \,.
\end{align}
Going here from left to right and starting from 0, the $j^{th}$ 
diagonal element of $D[n]$ is given by
\begin{equation}
\left(D[n]\right)_{j} = (-1)^{\lfloor \frac{j}{n} \rfloor}
\label{Dnj} \,.
\end{equation}
We now observe that $\sigma^{z}_{0} = D[1]$, and 
$\sigma^{z}_{l} = D[2^{l}]$. Hence, the $j^{th}$ diagonal element
of the operator whose expectation value we wish to compute is given 
by 
\begin{align}
\left(\sigma^{z}_{0}\, \sigma^{z}_{l}\right)_{j} &= 
\left(D[1]\right)_{j}\, \left(D[2^{l}]\right)_{j} \non \\
&= (-1)^{j}\, (-1)^{\lfloor \frac{j}{2^{l}} \rfloor} \,, \non \\
&= : \epsilon^{(l)}[j] \,, \qquad j = 0, \ldots, 2^{L}-1 \,,
\label{epsilonjdef}
\end{align}
where the products on the right-hand-side are ordinary products 
of scalars. We conclude that the coefficients in \eqref{corrmeas}
are given by 
\begin{equation}
c^{(l)}_{i_{L-1} \ldots i_{0}} = \epsilon^{(l)}[\sum_{k=0}^{L-1}2^{k} i_{k}]  
\label{cvalues}\,,
\end{equation}
where $\epsilon^{(l)}[j]$ is defined in \eqref{epsilonjdef}. As a 
simple example, for the case $L=2, l=1$, Eqs. \eqref{corrmeas} and \eqref{cvalues} give 
\eqref{warmup2}.

\subsection{Estimating the number of shots}

In order to perform shot-based measurements of the correlators \eqref{corrfunc}, how many shots should 
we use? In order to address this question, we begin by recalling a 
famous result associated with the Law of Large Numbers (see e.g. 
\cite{enwiki:1062022078}): let $\xi_{i}$ be independent, identically 
distributed random variables, which all have the same expected value 
${\rm E}[\xi_{i}] = 
\mu$ and variance ${\rm Var}[\xi_{i}] = \sigma^{2}$. Then a sample of $n$ 
such random variables has the average
\begin{equation}
\bar{\xi} = \frac{1}{n}\left(\xi_{1} + \ldots + \xi_{n} \right) \,,
\end{equation}
and the variance of the sample average is given by\footnote{The proof 
is short:
\begin{align}
{\rm Var}[\bar{\xi}] &=	{\rm Var}[\frac{1}{n}\left(\xi_{1} + \ldots + \xi_{n} \right)] \non \\
&=\frac{1}{n^{2}}{\rm Var}[\xi_{1} + \ldots + \xi_{n} ]\non \\
&=\frac{n\, {\rm Var}[\xi_{i}] }{n^{2}} = \frac{{\rm 
Var}[\xi_{i}]}{n}\,, \non
\end{align}
where we pass to the second line using the fact ${\rm Var}[a\, \xi] = 
a^{2} {\rm Var}[\xi]$, and we pass to the third line using the fact 
that the $\xi_{i}$ are independent.}
\begin{equation}
\epsilon^{2} := {\rm Var}[\bar{\xi}] = \frac{{\rm Var}[\xi_{i}]}{n}  \,.
\label{LLN}
\end{equation}
In the usual application to quantum computing, 
$\xi_{i}$ is regarded as an operator whose expectation value 
${\rm E}[\xi_{i}] = \langle \psi | \xi_{i} | \psi \rangle$ is 
measured by performing (multiple trials of) an experiment consisting 
of $n$ shots of a quantum circuit that prepares the state $| \psi \rangle$; 
the result \eqref{LLN} then relates the number of shots $n$ to the 
error $\epsilon$ of the measurement, see e.g. \cite{Wecker:2015}. 

However, our expectation value \eqref{corrfunc} is with respect to a state 
$|\psi_{0}\rangle$ that is prepared by the Bethe circuit with probability $|\alpha|^{2}$. 
Hence, the number of times $n$ that the state $|\psi_{0}\rangle$ is 
prepared is given by
\begin{equation}
n = N\, |\alpha|^{2} \,, 
\label{nNreltn}
\end{equation}
where $N$ is the number of shots of the Bethe circuit. Combining 
\eqref{LLN} and \eqref{nNreltn}, we see that the number of shots of 
the Bethe circuit is given by
\begin{equation}
N = \frac{{\rm Var}[\xi_{i}]}{|\alpha|^{2}\, \epsilon^{2}} \,, 	
\label{Nshots}
\end{equation}	
where here $\xi_{i} = \sigma^{z}_{0}\, \sigma^{z}_{l}$. We emphasize 
that $N$ is the number of shots required for the measurement of the 
full correlation function.

Let us now derive an upper bound on $N$. To this end, we observe that 
\begin{align}
{\rm Var}[\xi_{i}] &= {\rm E}[\xi_{i}^{2}] -{\rm E}[\xi_{i}]^{2} \non \\
&= \langle \psi_{0}| \left(\sigma^{z}_{0}\, \sigma^{z}_{l} \right)^{2}| \psi_{0} \rangle
- \left(\langle \psi_{0}| \sigma^{z}_{0}\, \sigma^{z}_{l} | \psi_{0} 
\rangle \right)^{2} \non \\
&= 1 - \left(\langle \psi_{0}| \sigma^{z}_{0}\, \sigma^{z}_{l} | \psi_{0} 
\rangle \right)^{2} \non \\
&\le 1 \,,
\label{varxi}
\end{align}
where we pass to the third line using the facts $\left(\sigma^{z}_{0}\, 
\sigma^{z}_{l} \right)^{2} = 1$ and $\langle \psi_{0}|  \psi_{0} 
\rangle =1$; and we pass to the last line using the fact  
$-1 \le \langle \psi_{0}| \sigma^{z}_{0}\, \sigma^{z}_{l} | \psi_{0} \rangle \le 1$.
It then follows from \eqref{Nshots} that
\begin{equation}
N \le N_{\rm max}\,, \qquad 	N_{\rm max} := \frac{1}{|\alpha|^{2}\, 
\epsilon^{2}} \,.
\label{Nmax}
\end{equation}
We note that $N_{\rm max}$ is independent of the value of $l$.
\footnote{It is also possible to derive a lower bound on $N$ in a 
similar way. Indeed, 
we see from Table \ref{table:corrfuncs} that the magnitudes of the correlators decrease with increasing
$L$ and $l$, the maximum occurring at $L=4$ and $l=1$:
\begin{equation}
|\langle \psi_{0}| \sigma^{z}_{0}\, \sigma^{z}_{l} | \psi_{0} \rangle| 
\le |\langle \psi_{0}| \sigma^{z}_{0}\, \sigma^{z}_{1} | 
\psi_{0}\rangle_{L=4}| = \frac{2}{3} \,. \non
\end{equation}
Using \eqref{varxi}, we obtain the inequality
\begin{equation}
{\rm Var}[\xi_{i}] 
= 1 - \left(\langle \psi_{0}| \sigma^{z}_{0}\, \sigma^{z}_{l} | \psi_{0} \rangle \right)^{2} 
\geq 1 - \left(\frac{2}{3}\right)^{2} = \frac{5}{9} \,. \non
\end{equation}
Recalling \eqref{Nshots}, we conclude that $N \geq N_\text{min}$, 
with $N_\text{min}=\frac{5}{9} N_\text{max}$. }

Summarizing: in order to measure the correlations \eqref{corrfunc} 
within error $\epsilon$, it suffices to perform (multiple trials of) an experiment consisting
of $N_{\rm max}$ shots of the Bethe circuit, where $N_{\rm max}$ is 
given by \eqref{Nmax}, and $|\alpha|^{2}$ is given by \eqref{successprobab}.

\subsection{Simulations for small $L$ values}

We checked our results \eqref{corrmeas}, \eqref{cvalues} and
\eqref{Nmax} by measuring the spin-spin correlation functions
\eqref{corrfunc} using the IBM Qiskit qasm simulator (without noise)
for small values of $L$.  Specifically, we set $\epsilon = 0.01$, and
we performed 100 trials (in order to accumulate
sufficient statistics to compute mean and standard deviation) of an
experiment consisting of $N$ shots of the Bethe circuit, with $N
\approx N_{\rm max}$ \eqref{Nmax}, and using \eqref{corrmeas}, \eqref{cvalues}  for the
measurements.  In the experimental (``exp'') columns of Table
\ref{table:corrfuncs}, the values of $|\alpha|^{2}$ and $N$ are noted,
and the mean and standard deviation of these 100 trials are reported.
We observe that the experimental standard deviations are all within
the specified error $\epsilon = 0.01$, thereby providing support for
\eqref{Nmax}.  The theoretical (``th'') columns of Table
\ref{table:corrfuncs} with $L=4, 6, 8$ display the values obtained
using {\tt Mathematica}.  For comparison, the results for $L=\infty$,
which are obtained from the literature, are also
displayed.\footnote{Remarkably, the correlation functions for
$L=\infty$ can be expressed as polynomials in $\ln 2$ and values of
the Riemann zeta function at odd arguments with rational coefficients
\cite{Boos:2005vu, Korepin:2006jvi}.}

\begin{table}[htb]
\footnotesize
\centering
\begin{tabular}{|c|cc|cc|cc|c|}
\hline
& $L=4$ & &  $L=6$ &  & $L=8 $ & & $L=\infty$ \\ 
\hline
$l$ & th & exp               & th & exp             & th & 
exp  & th \\ 
    &    & $|\alpha|^{2}=0.5$   &    & $|\alpha|^{2}=0.157232$ &    & $|\alpha|^{2}=0.0361418$ & \\
    &    & $N=2\times 10^{4}$   &    & $N=6.4\times 10^{4}$ &    & 
	$N=2.8\times 10^{5}$ & \\	
\hline
1 & -0.666667 & $-0.6657 \pm 0.0076$ &   -0.622839 & $-0.6230 \pm 
0.0072$ & -0.608516 & $-0.6091 \pm 0.0077$ & -0.590863 \cite{Hulthen:1938} \\
2 & 0.333333  & $0.3309 \pm 0.0090$ &   0.27735   & $0.2777 \pm 
0.0086$ &  0.261037 & $0.2602 \pm 0.0086$ & 0.242719 \cite{Takahashi:1977}\\
3 & --        & -- &  -0.309022  & $-0.3102 \pm 0.0090$ & -0.251937 
& $-0.2519 \pm 0.0088$ & -0.200995 \cite{Sakai:2003pc}\\
4 & --        & -- &    --       & -- & 0.198831  & $0.1988 \pm 
0.0093$ & 0.138611 \cite{Boos:2004sk}\\
\hline
\end{tabular}
\caption{The spin-spin correlation functions 
$\langle \psi_{0}| \sigma^{z}_{0}\, \sigma^{z}_{l}| \psi_{0} \rangle$, 
where $| \psi_{0} \rangle$ is the normalized ground state 
of the antiferromagnetic length-$L$ Hamiltonian \eqref{Hamiltonian}}
\label{table:corrfuncs}
\end{table}

\subsection{Larger $L$ values}\label{sec:long}

\begin{figure}[htb]
	\centering
	\includegraphics[width=0.5\hsize]{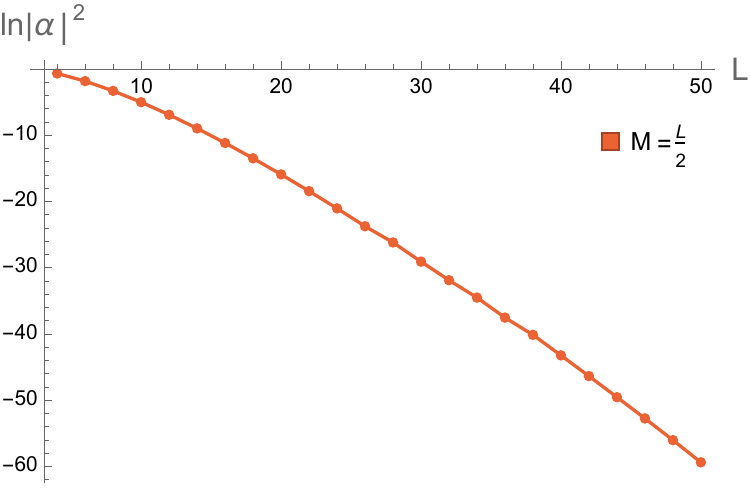}
	\caption{Logarithm of the success probability $\ln |\alpha|^{2}$ for the 
	antiferromagnetic ground state as a function of chain length $L$}
	\label{fig:successprobab}
\end{figure}

As can already be seen from Table \ref{table:corrfuncs}, as the length $L$
of the chain increases, the number of shots $N$ that are 
needed to measure the correlators \eqref{corrfunc} to within a 
specified (fixed) error $\epsilon$ increases. Indeed,
the success probability $|\alpha|^{2}$ decreases exponentially with 
$L$, as expected from \eqref{limit1} since 
$M=L/2$, and as shown in Fig. \ref{fig:successprobab}. 
Correspondingly, there is an exponential increase in $N_{\rm 
max}$. For example, for the moderate value $L=40$, we find
$|\alpha|^{2} \sim 5 \times 10^{-20}$, and therefore $N_{\rm max} \sim 
2 \times 10^{23}$, which is evidently impractical. 

As noted in \cite{VanDyke:2021kvq}, amplitude
amplification \cite{Brassard:2000} can generally be used to compensate for low
success probability. However, for the problem at hand,
the required number of iterations of amplitude amplification would render the circuit 
impractically deep. Indeed, after just one iteration, 
the success probability goes up from $\sin^2(\theta_{a}) := |\alpha|^{2}$ to 
$\sin^2(3\theta_{a})$ \cite{Brassard:2000}.
For $|\alpha|^{2}$ exponentially small, so that $\theta_{a} \approx 
|\alpha|$, this implies a 9-fold amplification, which 
is clearly insufficient.  The number $m$ of iterations needed to 
achieve success probability near 1 is given by
\begin{equation}
m = \left\lfloor \frac{\pi}{4 \theta_{a}} \right\rfloor \approx \frac{1}{\theta_{a}}
\approx \frac{1}{|\alpha|}  \,,
\end{equation}
see Theorem 2 in \cite{Brassard:2000}.
For the $L=40$ example considered above, $m\sim 10^9$. Since each iteration
effectively applies the Bethe circuit twice, the total circuit becomes
impractically deep. 

\section{Discussion}\label{sec:end}

We have found a simple exact formula \eqref{successprobab} for the
probability that the Bethe circuit \cite{VanDyke:2021kvq} successfully
prepares an eigenstate of the Heisenberg Hamiltonian
\eqref{Hamiltonian} corresponding to real Bethe roots.  
We have argued that, for large $L$ and fixed number of 
Bethe roots $M$, the success probability approaches $1/M!$ \eqref{limit1}.
We have also demonstrated the feasibility of using the Bethe circuit to
compute the spin-spin correlation functions \eqref{corrfunc} for 
small values of $L$, see Table \ref{table:corrfuncs}. 
However, we see from Fig. \ref{fig:successprobab} that the success probability decreases 
exponentially with $L$ for the antiferromagnetic ground state (for 
which $M=L/2$), which precludes the
computation of these correlation functions for moderate values of $L$.
We have considered here the optimal situation of no noise; of course,
the presence of noise would make matters worse.

A contribution to the exponential decrease of the success probability for increasing $M$ 
comes from the $1/\sqrt{M!}$ factors that are
associated with the permutation labels, as noted below
\eqref{statePsi}. Since the sum over permutations is an indispensable 
ingredient of the Bethe wavefunction \eqref{wavefunction}, we expect that 
any probabilistic algorithm for preparing Bethe states will 
necessarily involve creation of a superposition state of all 
permutations, such as \eqref{PLstate}; 
therefore, its success probability will necessarily have those $1/\sqrt{M!}$ 
factors.

The Bethe algorithm \cite{VanDyke:2021kvq}, which is for a closed spin
chain with periodic boundary conditions, has recently been extended to
the case of open spin chains \cite{VanDyke:2022}.  The latter
algorithm is also probabilistic.  We expect that the formula for
success probability \eqref{successprobab} can be generalized to the
open-chain case.  However, the above argument suggests that the success
probability for the open chain also decreases rapidly with $M$ 
(perhaps faster, since there is also a sum over the $2^{M}$ possible 
reflections), which would imply that -- as in the periodic case -- the correlation
functions \eqref{corrfunc} could be computed only for small values of $L$.

We have considered in Sec. \ref{sec:corr} an application of the Bethe
circuit involving antiferromagnetic {\em ground} states, which have the
maximum possible value of $M$ for a given value of $L$ (namely, $M=L/2$), and
which correspondingly have the smallest possible success
probability.  
Nevertheless, it could be feasible to prepare 
states with small values of $M$ -- even for moderate values of $L$ 
--  that would not be classically simulable, and which could still have
interesting applications \cite{VanDyke:2021kvq}. Indeed, as shown in 
Fig. \ref{fig:successprobabM1234}, the success probabilities for small 
values of $M$ are non-negligible, and are essentially independent of $L$.
Such states would correspond to high-energy (low-energy) {\em excited} 
states of the antiferromagnetic (ferromagnetic) Hamiltonian, 
respectively.

\section*{Acknowledgments} 
We thank John Van Dyke for helpful correspondence and comments on
a draft. We also thank Nikolai Kitanine for helpful correspondence.

\appendix

\section{Conjectures for estimates of the Gaudin \\
determinant}\label{sec:conjectures}

We note here some conjectures for bounds on the Gaudin 
determinant, which arose from our study of the success probability.

The physical requirement $|\alpha|^{2}\le 1$ together with \eqref{successprobab} 
imply the following upper bound on the Gaudin determinant
\begin{equation}
\det G \le \frac{L!\, M!}{(L-M)!} \,.
\label{conjecture0}
\end{equation}
It would be interesting to find a proof of this result (at least for 
the class of states considered here, namely, $M \in [1, L/2]$ and $L$ are
finite integers with $L$ even, and $k_{0}, \ldots, 
k_{M-1}$ are real and satisfy the Bethe equations) 
directly from the definition of the Gaudin matrix 
\eqref{Gaudinmatrix}. If one could show that the Gaudin matrix is 
positive (which we have verified for many examples), then 
the Perron-Frobenius theorem could be used to help show that
$\det G \le L^{M} \le \frac{L!\, M!}{(L-M)!}$. 

From an analysis of many states with real Bethe roots 
(all such states up to $L=20$, and selected cases up to $L=500$), we 
find that the success probability for {\em most} states 
satisfies the stronger bound\footnote{It was suggested in
\cite{VanDyke:2021kvq} that the ``worst-case'' success probability 
is $1/M!$, meaning $|\alpha|^{2}\ge 1/M!$, which we find is not satisfied 
for most cases. On the contrary, $1/M!$ appears to generally be the {\em best-case} 
success probability.}
\begin{equation}
|\alpha|^{2} \le \frac{1}{M!}  \qquad \text{(for most states)} \,,
\label{conjecture1}
\end{equation}
which is compatible with \eqref{limit1}, and correspondingly
\begin{equation}
\det G \le \frac{L!}{(L-M)!}  \qquad \text{(for most states)}   \,.
\label{conjecture2}
\end{equation}
However, we have identified some ``exceptional'' states that slightly violate these bounds.
These exceptional states are all characterized by sets of $M$ 
equally-spaced ``counting numbers'' 
\begin{equation}
\{-\frac{M-1}{2}, -\frac{M-1}{2} +1, \ldots, \frac{M-1}{2}-1\,, 
\frac{M-1}{2} \} \,,
\end{equation}
for certain values of $M\ge 3$ and $L\ge 22$, for 
which $|\alpha|^{2}\, M! = 1 + \delta$, where $\delta > 0$ is of order 
$10^{-5}$ for $L=22, M=3$, and reaches $10^{-2}$ for $L=500, M=75$. 
However, consistent with  \eqref{limit1}, we observe that $\delta \rightarrow 0$ for 
fixed $M$ and sufficiently large $L$.


\begin{thebibliography}{10}

\bibitem{Bethe:1931hc}
H.~Bethe, ``{On the theory of metals. 1. Eigenvalues and eigenfunctions for the
  linear atomic chain},''
\href{http://dx.doi.org/10.1007/BF01341708}{{\em Z. Phys.} {\bfseries 71}
  (1931) 205--226}.

\bibitem{Faddeev:1981ft}
L.~D. Faddeev and L.~A. Takhtajan, ``{Spectrum and scattering of excitations in
  the one-dimensional isotropic Heisenberg model},''
  \href{http://dx.doi.org/10.1007/BF01087245}{{\em Zap. Nauchn. Semin.}
  {\bfseries 109} (1981) 134--178}.

\bibitem{Gaudin:1983}
M.~Gaudin, {\em {La fonction d'onde de Bethe}}.
\newblock Masson, 1983.
\newblock English translation by J.-S. Caux, {\em The Bethe wavefunction}, CUP,
  2014.

\bibitem{VanDyke:2021kvq}
J.~S. Van~Dyke, G.~S. Barron, N.~J. Mayhall, E.~Barnes, and S.~E. Economou,
  ``{Preparing Bethe Ansatz Eigenstates on a Quantum Computer},'' {\em PRX
  Quantum} {\bfseries 2} (2021) 040329,
  \href{http://arxiv.org/abs/2103.13388}{{\ttfamily arXiv:2103.13388
  [quant-ph]}}.

\bibitem{Childs:2012}
A.~M. Childs and N.~Wiebe, ``{Hamiltonian simulation using linear combinations
  of unitary operations},'' {\em Quant. Inform. \& Comp.} {\bfseries 12}
  (November, 2012) 901, \href{http://arxiv.org/abs/1202.5822}{{\ttfamily
  arXiv:1202.5822 [quant-ph]}}.

\bibitem{Berry:2015}
D.~W. Berry, A.~M. Childs, R.~Cleve, R.~Kothari, and R.~D. Somma, ``Simulating
  {Hamiltonian} {Dynamics} with a {Truncated} {Taylor} {Series},'' {\em Phys.
  Rev. Lett.} {\bfseries 114} no.~9, (2015) 090502,
  \href{http://arxiv.org/abs/1412.4687}{{\ttfamily arXiv:1412.4687
  [quant-ph]}}.

\bibitem{Nepomechie:2021bethe}
R.~I. Nepomechie, ``{Bethe ansatz on a quantum computer?},'' {\em Quant. Inf.
  \& Comp.} {\bfseries 21} (2021) 255--265,
  \href{http://arxiv.org/abs/2010.01609}{{\ttfamily arXiv:2010.01609
  [quant-ph]}}.

\bibitem{Gaudin:1981cyg}
M.~Gaudin, B.~M. McCoy, and T.~T. Wu, ``{Normalization sum for the Bethe's
  hypothesis wave functions of the Heisenberg-Ising chain},''
  \href{http://dx.doi.org/10.1103/PhysRevD.23.417}{{\em Phys. Rev. D}
  {\bfseries 23} no.~2, (1981) 417}.

\bibitem{Korepin:1982gg}
V.~E. Korepin, ``{Calculation of norms of Bethe wave functions},''
  \href{http://dx.doi.org/10.1007/BF01212176}{{\em Commun. Math. Phys.}
  {\bfseries 86} (1982) 391--418}.

\bibitem{Hulthen:1938}
L.~Hulth\'en, ``{\"Uber das Austauschproblem eines Kristalles},'' {\em Ark.
  Math. Astron. Fys.} {\bfseries 26A} (1938) .

\bibitem{Luther:1975wr}
A.~Luther and I.~Peschel, ``{Calculation of critical exponents in
  two-dimensions from quantum field theory in one-dimension},''
  \href{http://dx.doi.org/10.1103/PhysRevB.12.3908}{{\em Phys. Rev. B}
  {\bfseries 12} (1975) 3908--3917}.

\bibitem{Takahashi:1977}
M.~Takahashi, ``{Half-filled Hubbard model at low temperature},'' {\em J.
  Phys.} {\bfseries C 10} (1977) 1289.

\bibitem{Lin:1991}
H.~Q. Lin and D.~K. Cambell, ``{Spin-spin correlations in the one-dimensional
  spin-1/2, antiferromagnetic Heisenberg chain},'' {\em J. Applied Phys.}
  {\bfseries 69} (1991) 5947.

\bibitem{Hallberg:1995}
K.~A. Hallberg, P.~Horsch, and G.~Martínez, ``{Numerical renormalization-group
  study of the correlation functions of the antiferromagnetic spin-1/2
  Heisenberg chain},'' {\em Phys. Rev. B} {\bfseries 52} no.~2, (1995)
  R719–R722, \href{http://arxiv.org/abs/cond-mat/9505132}{{\ttfamily
  arXiv:cond-mat/9505132 [cond-mat]}}.

\bibitem{Kitanine:2000srg}
N.~Kitanine, J.~M. Maillet, and V.~Terras, ``{Correlation functions of the XXZ
  Heisenberg spin-$\frac1 2$ chain in a magnetic field},''
  \href{http://dx.doi.org/10.1016/S0550-3213(99)00619-7}{{\em Nucl. Phys. B}
  {\bfseries 567} (2000) 554--582},
  \href{http://arxiv.org/abs/math-ph/9907019}{{\ttfamily
  arXiv:math-ph/9907019}}.

\bibitem{Kitanine:2002mn}
N.~Kitanine, J.~M. Maillet, N.~A. Slavnov, and V.~Terras, ``{Spin spin
  correlation functions of the XXZ - 1/2 Heisenberg chain in a magnetic
  field},'' \href{http://dx.doi.org/10.1016/S0550-3213(02)00583-7}{{\em Nucl.
  Phys. B} {\bfseries 641} (2002) 487--518},
  \href{http://arxiv.org/abs/hep-th/0201045}{{\ttfamily arXiv:hep-th/0201045}}.

\bibitem{Lukyanov:2002fg}
S.~L. Lukyanov and V.~Terras, ``{Long distance asymptotics of spin spin
  correlation functions for the XXZ spin chain},''
  \href{http://dx.doi.org/10.1016/S0550-3213(02)01141-0}{{\em Nucl. Phys. B}
  {\bfseries 654} (2003) 323--356},
  \href{http://arxiv.org/abs/hep-th/0206093}{{\ttfamily arXiv:hep-th/0206093}}.

\bibitem{Sakai:2003pc}
K.~Sakai, M.~Shiroishi, Y.~Nishiyama, and M.~Takahashi, ``{Third neighbor
  correlators of spin 1/2 Heisenberg antiferromagnet},''
  \href{http://dx.doi.org/10.1103/PhysRevE.67.065101}{{\em Phys. Rev. E}
  {\bfseries 67} (2003) 065101},
  \href{http://arxiv.org/abs/cond-mat/0302564}{{\ttfamily
  arXiv:cond-mat/0302564}}.

\bibitem{Boos:2004sk}
H.~E. Boos, M.~Shiroishi, and M.~Takahashi, ``{First principle approach to
  correlation functions of spin-1/2 Heisenberg chain: Fourth-neighbor
  correlators},'' \href{http://dx.doi.org/10.1016/j.nuclphysb.2005.01.041}{{\em
  Nucl. Phys. B} {\bfseries 712} (2005) 573--599},
  \href{http://arxiv.org/abs/hep-th/0410039}{{\ttfamily arXiv:hep-th/0410039}}.

\bibitem{Boos:2005vu}
H.~Boos, M.~Jimbo, T.~Miwa, F.~Smirnov, and Y.~Takeyama, ``{Density matrix of a
  finite sub-chain of the Heisenberg anti-ferromagnet},''
  \href{http://dx.doi.org/10.1007/s11005-006-0054-x}{{\em Lett. Math. Phys.}
  {\bfseries 75} (2006) 201--208},
  \href{http://arxiv.org/abs/hep-th/0506171}{{\ttfamily arXiv:hep-th/0506171}}.

\bibitem{Korepin:2006jvi}
V.~E. Korepin and O.~I. Patu, ``{XXX spin chain: From Bethe solution to open
  problems},'' \href{http://dx.doi.org/10.22323/1.038.0006}{{\em PoS}
  {\bfseries SOLVAY} (2006) 006},
  \href{http://arxiv.org/abs/cond-mat/0701491}{{\ttfamily
  arXiv:cond-mat/0701491}}.

\bibitem{Caux:2005a}
J.-S. {Caux} and J.~M. {Maillet}, ``{Computation of Dynamical Correlation
  Functions of Heisenberg Chains in a Magnetic Field},''
  \href{http://dx.doi.org/10.1103/PhysRevLett.95.077201}{{\em Phys. Rev. Lett.}
  {\bfseries 95} no.~7, (2005) 077201},
  \href{http://arxiv.org/abs/cond-mat/0502365}{{\ttfamily
  arXiv:cond-mat/0502365 [cond-mat.str-el]}}.

\bibitem{Caux:2005b}
J.-S. {Caux}, R.~{Hagemans}, and J.~M. {Maillet}, ``{Computation of dynamical
  correlation functions of Heisenberg chains: the gapless anisotropic
  regime},'' \href{http://dx.doi.org/10.1088/1742-5468/2005/09/P09003}{{\em J.
  Stat. Mech.} {\bfseries 2005} no.~9, (2005) 09003},
  \href{http://arxiv.org/abs/cond-mat/0506698}{{\ttfamily
  arXiv:cond-mat/0506698 [cond-mat.str-el]}}.

\bibitem{Caux:2009}
J.-S. {Caux}, ``{Correlation functions of integrable models: A description of
  the ABACUS algorithm},'' \href{http://dx.doi.org/10.1063/1.3216474}{{\em J.
  Math. Phys.} {\bfseries 50} no.~9, (2009) 095214--095214},
  \href{http://arxiv.org/abs/0908.1660}{{\ttfamily arXiv:0908.1660
  [cond-mat.str-el]}}.

\bibitem{Somma:2002}
R.~Somma, G.~Ortiz, J.~E. Gubernatis, E.~Knill, and R.~Laflamme, ``{Simulating
  physical phenomena by quantum networks},'' {\em Phys. Rev.} {\bfseries A 65}
  no.~4, (2002) , \href{http://arxiv.org/abs/0108146}{{\ttfamily arXiv:0108146
  [quant-ph]}}.

\bibitem{Wecker:2015}
D.~Wecker, M.~B. Hastings, N.~Wiebe, B.~K. Clark, C.~Nayak, and M.~Troyer,
  ``{Solving strongly correlated electron models on a quantum computer},'' {\em
  Phys. Rev.} {\bfseries A 92} no.~6, (2015) ,
  \href{http://arxiv.org/abs/1506.05135}{{\ttfamily arXiv:1506.05135
  [quant-ph]}}.

\bibitem{enwiki:1062022078}
{Wikipedia contributors}, ``Law of large numbers --- {Wikipedia}{,} the free
  encyclopedia.''
  \url{https://en.wikipedia.org/w/index.php?title=Law_of_large_numbers&oldid=1062022078},
  2021.
\newblock [Online; accessed 27-December-2021].

\bibitem{Brassard:2000}
G.~Brassard, P.~Høyer, M.~Mosca, and A.~Tapp, ``{Quantum amplitude
  amplification and estimation},'' {\em Quant. Comp. \& Inform.} (2002)
  53–74, \href{http://arxiv.org/abs/quant-ph/0005055}{{\ttfamily
  arXiv:quant-ph/0005055 [quant-ph]}}.

\bibitem{VanDyke:2022}
J.~S. Van~Dyke, E.~Barnes, S.~E. Economou, and R.~I. Nepomechie, ``{Preparing
  exact eigenstates of the open XXZ chain on a quantum computer},''
  \href{http://arxiv.org/abs/2109.05607}{{\ttfamily arXiv:2109.05607
  [quant-ph]}}.

\end{thebibliography}

\providecommand{\href}[2]{#2}\begingroup\raggedright\endgroup

\end{document}